\documentclass[sn-mathphys,Numbered]{sn-jnl}

\usepackage{graphicx}%
\usepackage{multirow}%
\usepackage{amsmath,amssymb,amsfonts}%
\usepackage{amsthm}%
\usepackage{mathrsfs}%
\usepackage[title]{appendix}%
\usepackage{xcolor}%
\usepackage{textcomp}%
\usepackage{manyfoot}%
\usepackage{booktabs}%
\usepackage{algorithm}%
\usepackage{algorithmicx}%
\usepackage{algpseudocode}%
\usepackage{listings}%

\usepackage[normalem]{ulem}

\theoremstyle{thmstyleone}%
\theoremstyle{thmstyletwo}%

\theoremstyle{thmstylethree}%

\newcommand{\ulmo}{{\sc Ulmo}}
\newcommand{\enki}{{\sc Enki}}

\newcommand{\vitmae}{ViTMAE}
\newcommand{\tper}{\ensuremath{t_\%}}
\newcommand{\mper}{\ensuremath{m_\%}}

\newcommand{\tbest}{\ensuremath{20}}  

\newcommand{\cutoutsz}{\ensuremath{64\times 64 \, \rm pixels}}

\newcommand{\patchsz}{\ensuremath{4\times 4 \, \rm pixels}}
\newcommand{\floor}{0.014}
\newcommand{\pscale}{9.1}

\newcommand{\sst}{SST}
\newcommand{\ssta}{SSTa}

\newcommand{\llc}{LLC4320}
\newcommand{\ntrain}{2,623,152}
\newcommand{\nvalid}{655,788}

\newcommand{\nviirs}{923,751}
\newcommand{\inpaintname}{biharmonic}
\newcommand{\rmsev}{\ensuremath{{\rm RMSE}_{\rm VIIRS}}}

\newcommand{\errav}{\ensuremath{\epsilon_{\rm AVHRR}}}
\newcommand{\errviirs}{\ensuremath{\epsilon_{\rm VIIRS}}}

\newcommand{\stdT}{\ensuremath{\sigma_T}}
\newcommand{\llulmo}{\ensuremath{LL_{\rm Ulmo}}}
\newcommand{\rmse}{\ensuremath{{\rm RMSE}}}
\newcommand{\renki}{\ensuremath{\rmse_{\rm Enki}}}
\newcommand{\rinpaint}{\ensuremath{\rmse_{\rm inpaint}}}

\newcommand{\snoin}{\ensuremath{\sigma_{noise\_o\_in}}}
\newcommand{\snoout}{\ensuremath{\sigma_{noise\_o\_out}}}

\begin{document}

\title{Mitigating masked pixels in climate-critical datasets}



\author*[1]{\fnm{Angelina} \sur{Agabin}}\email{aagabin@ucsc.edu}
\equalcont{These authors contributed equally to this work.}

\author[2,3]{\fnm{J. Xavier} \sur{Prochaska}}\email{jxp@ucsc.edu}
\equalcont{These authors contributed equally to this work.}

\author[4]{\fnm{Peter} \sur{Cornillon}}\email{pcornillon@uri.edu}

\author[5]{\fnm{Christian~E.} \sur{Buckingham}}\email{christian.buckingham@umassd.edu}

\affil*[1]{\orgdiv{Department of Applied Math}, 
\orgname{University of California, Santa Cruz},
\orgaddress{\street{1156 High St.}, 
\city{Santa Cruz}, 
\postcode{95064}, 
\state{CA}, 
\country{USA}}}

\affil*[2]{\orgdiv{Department of Ocean Sciences},
\orgname{University of California, Santa Cruz},
\orgaddress{\street{1156 High St.}, 
\city{Santa Cruz}, 
\postcode{95064}, 
\state{CA}, 
\country{USA}}}

\affil*[3]{\orgdiv{Department of Astronomy \& Astrophysics},
\orgname{University of California, Santa Cruz},
\orgaddress{\street{1156 High St.}, 
\city{Santa Cruz}, 
\postcode{95064}, 
\state{CA}, 
\country{USA}}}

\affil[4]{\orgdiv{Graduate School of Oceanography}, \orgname{University of Rhode Island}, \orgaddress{\street{215 South Ferry Road}, \city{Narragansett}, \postcode{02874}, \state{Rhode Island}, \country{USA}}}

\affil[5]{\orgdiv{School for Marine Science and Technology}, \orgname{University of Massachusetts, Dartmouth}, \orgaddress{\street{836 South Rodney French Blvd.}, \city{New Bedford}, \postcode{02747}, \state{Massachusetts}, \country{USA}}}


\abstract{
Remote sensing observations of the Earth's
surface are frequently stymied by clouds, water vapour, 
and aerosols in our atmosphere.
These degrade or preclude the measurement
of quantities critical to scientific and, hence, 
societal applications. 
In this study, we train a natural language processing (NLP)
algorithm with high-fidelity ocean simulations in order to accurately reconstruct masked or missing
data 
in sea surface temperature (SST)--i.e.~one of 54 essential climate variables identified by the Global Climate Observing System. 
We demonstrate that the \enki\ model repeatedly outperforms previously adopted
inpainting techniques by up to an order-of-magnitude
in reconstruction error, while displaying high performance even in circumstances where the majority
of pixels are masked.
Furthermore, experiments on real infrared sensor
data with masking fractions of at least 40\%\
show reconstruction errors 
of less than the known sensor uncertainty (RMSE~$< \sim 0.1$\,K). 
We attribute \enki's success to the attentive nature
of NLP combined with realistic \sst\ model outputs, an approach
that may be extended to other remote sensing variables.
This study demonstrates that systems built upon \enki\--or other advanced systems like it--may therefore yield the
optimal solution to
accurate estimates of otherwise missing or masked parameters
in climate-critical datasets sampling a rapidly changing Earth.
}


\keywords{Sea Surface Temperature, Machine Learning, Masked Autoencoder, Inpainting}



\maketitle

\section{Introduction}\label{sec1}

One of the most powerful means to assess fundamental properties of Earth 
is via remote sensing, satellite-bourne observations of its atmosphere,
land, and ocean surface.
Since the launch of 
the Television InfraRed Observation Satellite (TIROS) in 1960, the first of the non-military `weather satellites',
remote-sensing satellites have offered daily coverage of the
globe to monitor our atmosphere. 
In 1978, the launch of three satellites
directed the attention to our oceans, 
with sensors observing in the visible portion of the electromagnetic
(EM) spectrum to measure ocean color for biological applications, 
in the infrared (IR) range to estimate sea surface temperature
(SST), and in the microwave to estimate wind speed, 
sea surface height and \sst.
These programs were followed by a large number of internationally launched satellites carrying a broad range of sensors providing improved spatial, temporal and radiometric resolution for terrestrial, oceanographic, meteorological and cryospheric applications.

All satellite-borne sensors observing Earth's surface or atmosphere sample some portion 
of the EM spectrum, with the associated EM waves passing through some or all of the atmosphere. 
The degree to which the signal sampled is affected by the atmosphere is a strong function of the EM wavelength as well as the composition of the atmosphere, with wavelengths from the visible through the thermal infrared (400 nm -- 15$\mu$m) being the most affected. This is also the portion of the spectrum used to sample a wide range of surface parameters, such as land use, vegetation, ocean color, 
SST, snow cover, etc. Retrieval algorithms are designed to compensate for the atmosphere for many of these parameters,
but these algorithms fail if the density of particulates, such as dust or liquid or crystalin water--think clouds--is too large. The pixels for which this occurs are generally flagged and ignored.
This results in a gappy field, with the 
masked regions ranging from single pixels to regions covering tens of thousands of pixels in size. 
On average, for example, only
$\sim 15$\% of ocean pixels return an acceptable estimate of SST. 

This presents difficulties in the analysis of these fields, especially those requiring complete fields. 
To address such gaps, data from several different sources are often used with objective analysis interpolation programs to produce what are referred to as Level-4 products: gap-free fields (e.g., \cite{Reynolds2007}). 
Researchers have also introduced a diversity of algorithms to fill in clouds
(see \cite{reconstruct_review} for a review) 
including  methods using 
interpolation \cite{ulmo},
principal component analyses \cite{EOF2003,DINEOF2005},
and, most recently, convolutional neural networks \cite{DINCAE,larson2023}.
For \sst, these methods achieve average root mean square 
errors of $\approx 0.2-0.5$\,K and have input requirements
ranging from individual images to an extensive time series.

In this manuscript, we introduce a novel approach, 
inspired by the vision transformer masked autoencoder
(\vitmae) model of \cite{vitmae} 
to reconstruct masked pixels in satellite-derived fields.
Guided by the intuition that
  (1) natural images (e.g.\ dogs, landscapes)
  can be described by a language and therefore analyzed
  with natural language processing (NLP) techniques
  and
  (2) one can frequently recover the sentiment 
  of sentences that are missing words and then predict
  these words,
\cite{vitmae} demonstrated the remarkable effectiveness of 
\vitmae\ to reconstruct masked images.
This included images with 75\%\ masked data,
a remarkable inference.
Central to \vitmae's success was its training on a large corpus
of unmasked, natural images.
And, given the reduced complexity of most remote-sensing data compared to 
natural images, one may expect even better performance. 

In this study, we use the fine-scale ($1/48^\circ$, 90-level) ocean simulation
from the Estimating the Circulation and Climate of the Ocean (ECCO) project and referred to as {\llc} to train an implementation of the
\vitmae. 
The \llc\ model simulates $\approx 14$~months of the global ocean at a spatial resolution of $\approx 1-2$\,km with an hourly snapshot of model variables. 
Its standard output includes SST, and a comparison of the model
output with remote sensing data indicates excellent fidelity
\cite{ulmo_on_llc}. 
We thereby construct \enki, a \vitmae\ model trained on 
\sst\ ocean model outputs that may then be applied to 
actual remote sensing data.

We show below that {\enki} reconstructs images of 
SST anomalies (SSTa) 
far more accurately than conventional inpainting algorithms.
We demonstrate that the combination of
high-fidelty model output with  
state-of-the-art artificial intelligence
produces the unprecedented ability
to predict critical missing data
for both climate-critical and commercial applications.
Furthermore, the methodology allows for the comprehensive estimation
of uncertainty and an assessment of 
systematics.
The combined power of NLP algorithms and realistic model outputs
represent a terrific advance for image reconstruction of 
remote sensing applications.

\begin{figure}[h]%
\centering
\includegraphics[width=1.0\textwidth]{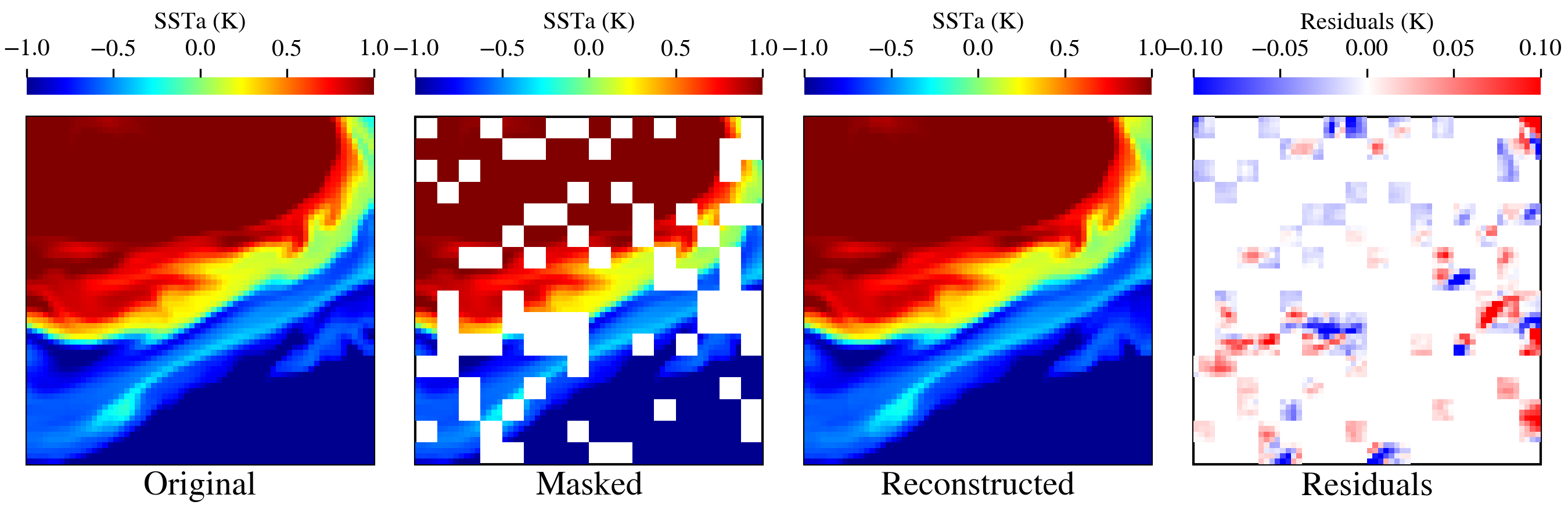}
\caption{
{\bf An example of the {\enki} model applied to simulated data with masked or missing pixels.} 
In this case, {\enki} was trained on realistic ocean model output consisting of $20$\% of pixels 
flagged as ``poor'' (masked) and applied to an image 
with $30$\% of pixels flagged as masked.
The panels are:
(a) original sea surface temperature anomalies (SSTa), 
(b)~masked SSTa in \patchsz\ patches covering 20\%\ of the image, 
(c)~reconstructed SSTa, and 
(d)~residual (reconstruction minus truth). 
Ignoring the image boundary (see Figure~\ref{fig:patches}),
the maximum reconstruction error is
only $\approx 0.19$\,K and the highest RMSE
in a single patch is $\approx 0.07$\,K with
an average RMSE~$\approx 0.02$\,K.
}
\label{fig:qualitative}
\end{figure}

\begin{figure}[h]%
\centering
\includegraphics[width=0.9\textwidth]{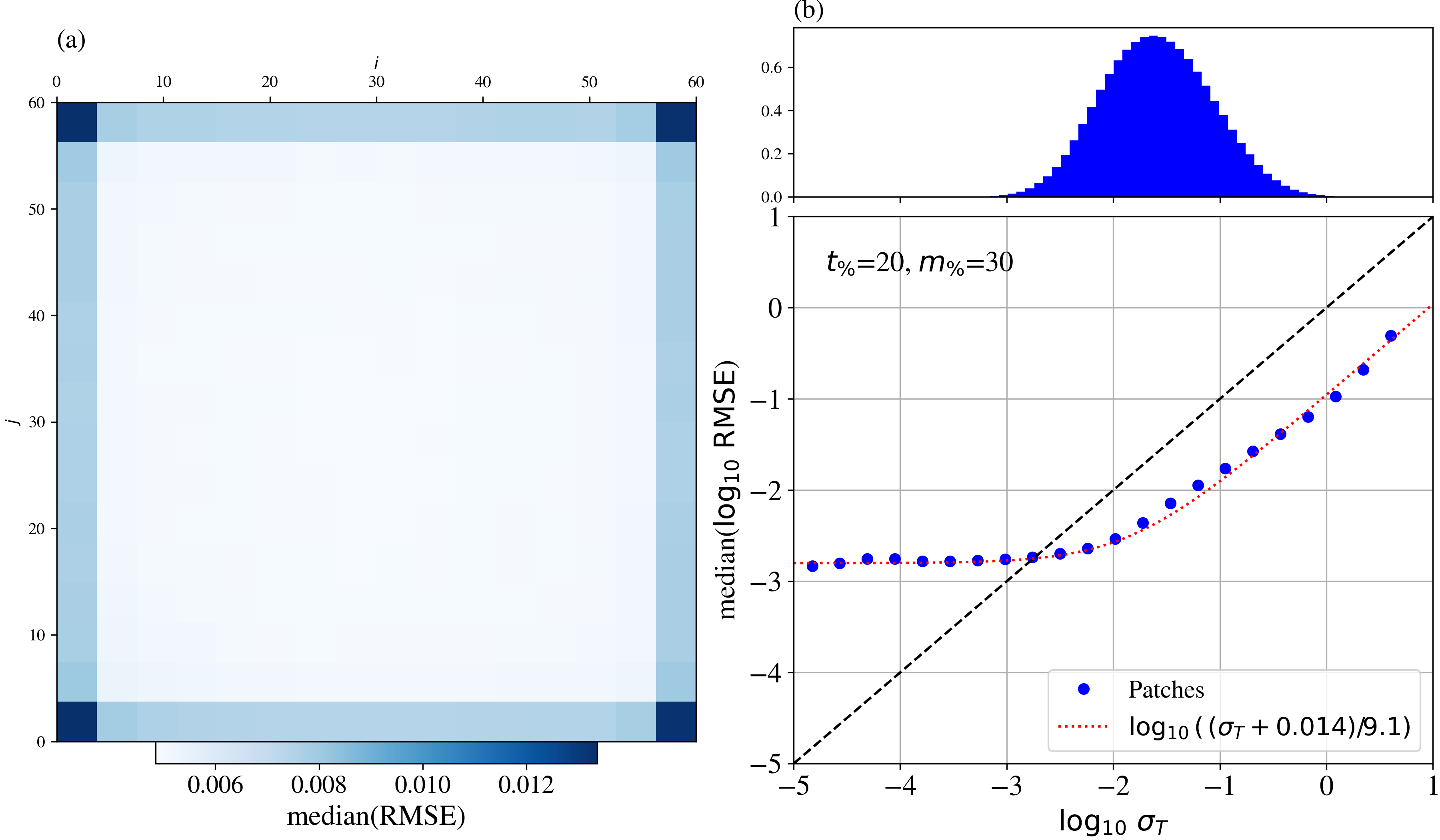}
\caption{ {\bf Results for individual \patchsz\ patches.}
Both panels present reconstruction results for the validation
dataset using a masking percentile of \mper=30 and the \tper=20 {\enki} model.
(a) Median \rmse\ as a function of the patch spatial location where
$i,j$ refers to the position of the lower-left corner of the patch.
Patches on the image boundary exhibit systematically higher \rmse\ 
and we advocate ignoring these in any image reconstruction application.
(b) Median of $\log_{10} \rmse$ for patches as a function of
$\log_{10}$ of the standard deviation of SST (\stdT) in the patch.
For patches with non-negligible structure $(\stdT > 10^{-2}$\,K),
the reconstruction \rmse\ is $\approx 10\times$ lower than random
(as described by the dashed one-to-one line).
The red curve is a two-parameter fit to the data. 
} 
\label{fig:patches}
\end{figure}

\begin{figure}[h]%
\centering
\includegraphics[width=0.9\textwidth]{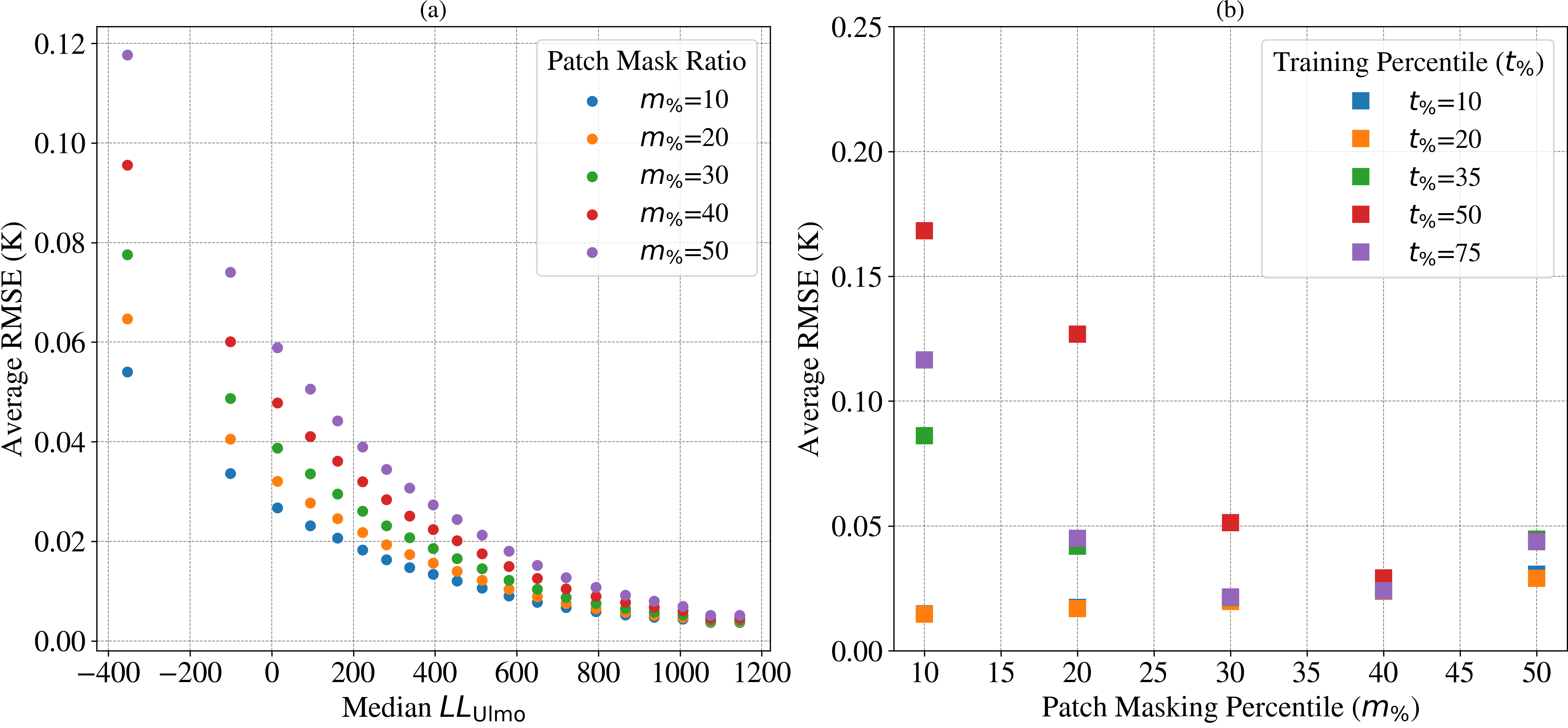}
\caption{
{\bf {\enki}'s reconstruction performance on the
validation dataset.}
(a) Model performance as a function of quantiles of
image complexity,
\llulmo, with higher values indicating lower complexity.
The reconstruction error is sensitive to the degree
of structure in the field, although for masking
percentile $\mper < 20$ even the most complex images
have $\rmse \lesssim 0.06$\,K.
(b) Model performance as a function of masking
percentile for the full set of trained {\enki} models.
Contrary to expectation, we find the \tper=\tbest\
model outperforms all others at all \mper.
The poorer performance of the $\tper > 20$ models at
low \mper\ indicates those models have not
learned the small-scale features present
in SST data.
}
\label{fig:cutouts}
\end{figure}

\section{Results}
\label{sec:results}

Figure~\ref{fig:qualitative} shows the reconstruction
of a representative
example from the validation dataset for the 
{\enki} model.  We refer to the image as a ``cutout''; it is a 
\cutoutsz\ section of the ocean model output with a scale of 
$\approx 144 \times 144 \, \rm km^2$.
In this case, the model was trained on 
cutouts characterized by $20$\% of the pixels being masked 
(\tper=20 model) but applied to a cutout having 30\% of its pixels 
masked (\mper=30).
Aside from patches along the outer edge of the input image,
it is difficult to visually differentiate the reconstructed pixels from the 
surrounding {\ssta} values. 
The greatest difference is 0.19~K and the highest RMSE in a
single $4\times4$ pixel patch is $\approx 0.07$\,K
with an average RMSE of $\approx 0.02$\,K.
As described below, 
the performance does degrade with higher
\mper\ and/or greater image complexity,
but to levels generally less than
standard sensor error.

Quantitatively, we consider first the 
model performance for individual
patches.  
Figure~\ref{fig:patches} presents the results 
of two analyses: 
  (a) the RMSE of reconstruction as a function of
  the patch spatial location in the image; 
  and
  (b) the quality of reconstruction as a function of
  patch complexity, defined by the standard deviation (\stdT)
  of the patch in the original image.
On the first point, it is evident from Figure~\ref{fig:patches}a
that {\enki} struggles
to faithfully reproduce data in patches on the image
boundary.  This is a natural outcome driven by the
absence of data on one or more sides of the patch
(i.e., partial extrapolation vs.\ interpolation).
Within this manuscript, we do not include boundary pixels or patches in any quantitative evaluation, 
and we emphasize that any systems built on a model like
{\enki} should ignore the boundary pixels in the reconstruction. 

Figure~\ref{fig:patches}b, meanwhile,
demonstrates that {\enki} reconstructs the data with an
RMSE that is over one order of magnitude
smaller than that anticipated from random chance.
Instead, the results track the relation
RMSE~$\approx (\stdT + \floor)/\pscale$.
We speculate that the ``floor'' in RMSE 
at $\stdT < 10^{-2}$\,K arises because
of the loss of information in tokenizing the 
image patches.
The nearly linear relation at larger \stdT, however, 
indicates that the model performance is fractionally
invariant with data complexity.
We examine these results further in Appendix~\ref{app:patches}.


Turning to performance at the full cutout level,
Figure~\ref{fig:cutouts}a shows results 
as a function of cutout complexity.
For the latter, we adopt a deep-learning
metric developed by \cite{ulmo} to identify 
outliers in \ssta\ imagery.
Their algorithm, named \ulmo, calculates a
log-likelihood value \llulmo\ designed to 
assess the probability of a given image 
occurring within a very large dataset of
SST images.
\cite{ulmo} and \cite{nenya}
demonstrate that data with the lowest \llulmo\ 
exhibit greater complexity, both in terms
of peak-to-peak temperature anomalies and also
in terms of the frequency and strength of fronts, etc.

Figure~\ref{fig:cutouts}a reveals that the 
reconstruction performance depends on \llulmo,
with the most complex cutouts showing 
$\rmse \approx 0.05-0.1$\,K.
For less complex data ($\llulmo > 200$),
the average $\rmse < 0.04$\,K which is effectively
negligible for most applications.
Even the largest {\rmse}s are smaller than 
the sensor errors
found by \cite{rs9090877} for the pixel-to-pixel noise in {\sst} fields retrieved from the Advanced Very High Resolution Radiometer 
(AVHRR; $\errav \lesssim 0.2$\,K), 
and comparable or better than
for the Visible-Infrared Imager-Radiometer Suite 
(VIIRS; $\errviirs \lesssim 0.1$\,K;
\cite{rs9090877}).

As described in Section~\ref{sec:methods}, we trained {\enki} 
with a range of training mask percentiles
expecting best performance with \tper=75 as
adopted by \cite{vitmae}.
Figure~\ref{fig:cutouts}b shows that for effectively
all masking percentiles \mper, the \tper=\tbest\ 
{\enki} model provides best performance.
We hypothesize that lower \tper\ models
are optimal for data with lower complexity
compared to natural images;
i.e., one can sufficiently generalize
with $\tper \ll 75\%$.
Furthermore, it is evident that models with
$\tper > 20$ have not learned the small-scale
structure apparent in SST imagery.

\begin{figure}[h]%
\centering
\includegraphics[width=0.9\textwidth]{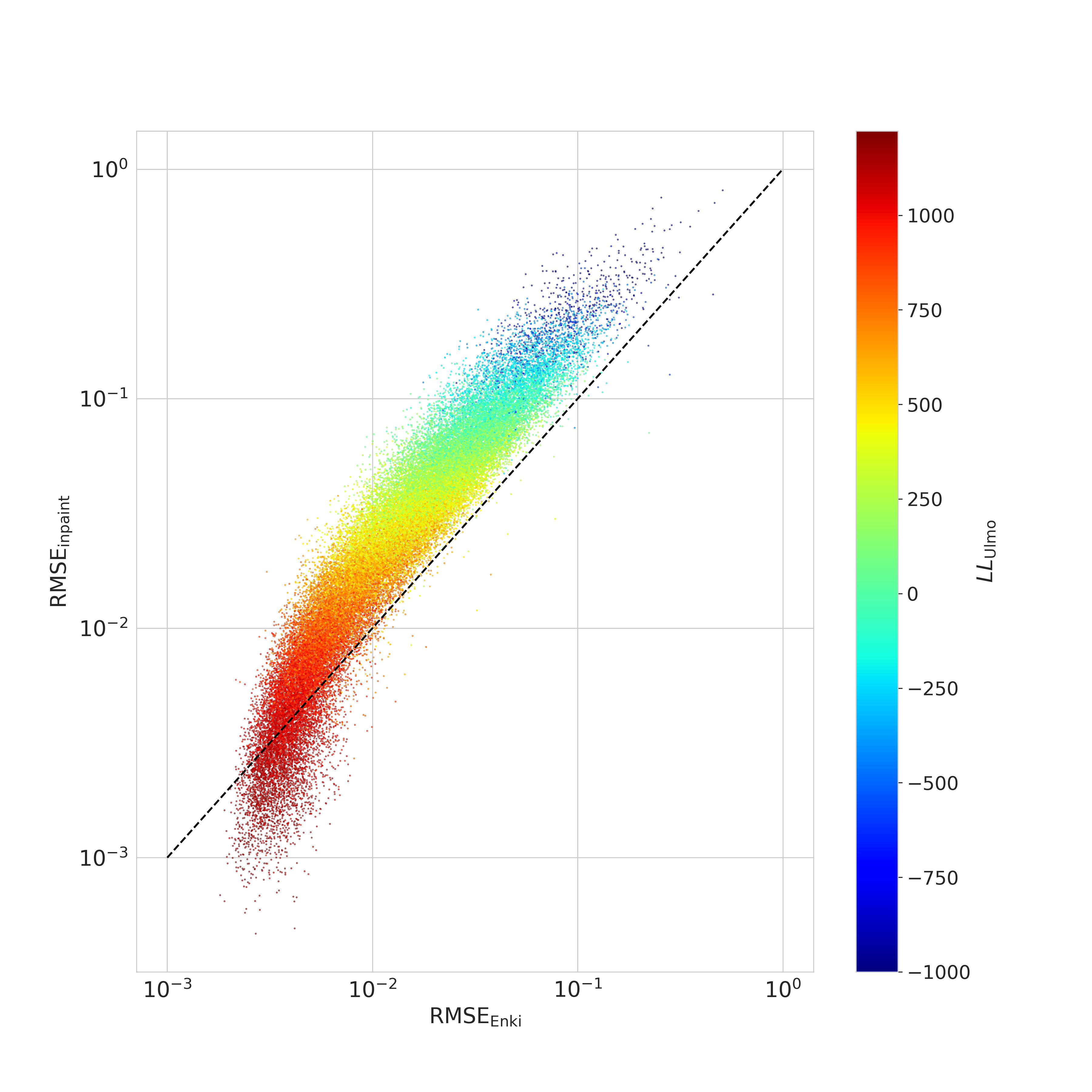}
\caption{
{\bf Comparison of reconstruction error for 
{\enki} (\renki) with the biharmonic inpainting
algorithm (\rinpaint) on the validation dataset.
} Colors denote the \llulmo\ metric \cite{ulmo}, which is a measure of image complexity; lower
values indicate higher image complexity.
In $>99.9\%$ of the cases, \enki\ outperforms the inpainting
algorithm, and often by more than an order-of-magnitude. This relationship holds 
independent of the image complexity. 
The results here correspond to the 
\tper=20 model applied to data with \mper=30 masking.
}
\label{fig:llc_inpainting}
\end{figure}

For a benchmark test, we compare the results from
{\enki} against the \inpaintname\ inpainting algorithm
(Figure~\ref{fig:llc_inpainting}).
This algorithm was selected because 
tests have indicated it faithfully reproduces sharp
gradients in SST \cite{ulmo}.
{\enki} outperforms that algorithm by an average factor
of $\approx 3\times$ for nearly all of the
cutouts except those with negligible structure and
very small RMSE ($< 0.01$\,K).
These results hold independent of image properties, e.g., 
image complexity.
Moreover, we found similar
results hold when comparing with other standard inpainting schemes 
(see Appendix~\ref{app:other}).

We have performed a similar benchmark comparison between \enki\ and 
other inpainting algorithms.  Appendix~\ref{app:other} describes
results for other interpolation schema with results similar (or better,
for \enki) to those presented in Figure~\ref{fig:llc_inpainting}.
In addition, we performed an experiment with one of the most frequently
adopted reconstruction techniques for \sst\ observations:
the DINEOF algorithm which predicts missing data based on empirical
orthogonal function fits to a sequence of masked observations \cite{DINEOF2005}.
As detailed in Appendix~\ref{app:other}, we applied DINEOF
to a 180-day sequence of \llc\ model output with the identical masking 
implemented in \enki\ (\mper=30).
For an example cutout of $\approx 144 \times 144 \, \rm km^2$
in the China Sea, DINEOF yield an average RMSE of 
$\approx 0.25$\,K (ignoring the boundary)
consistent with published results \cite{DINEOF2005,DINCAE}.
In contrast, we find \enki\ acheives an average RMSE
of $\approx 0.05$\,K, i.e.\ a five times improvement.
And, we emphasize these are "one-shot" reconstructions
and we have not trained \enki\ on a time series nor
with geographically limited regions.

\begin{figure}[h]%
\centering
\includegraphics[width=0.9\textwidth]{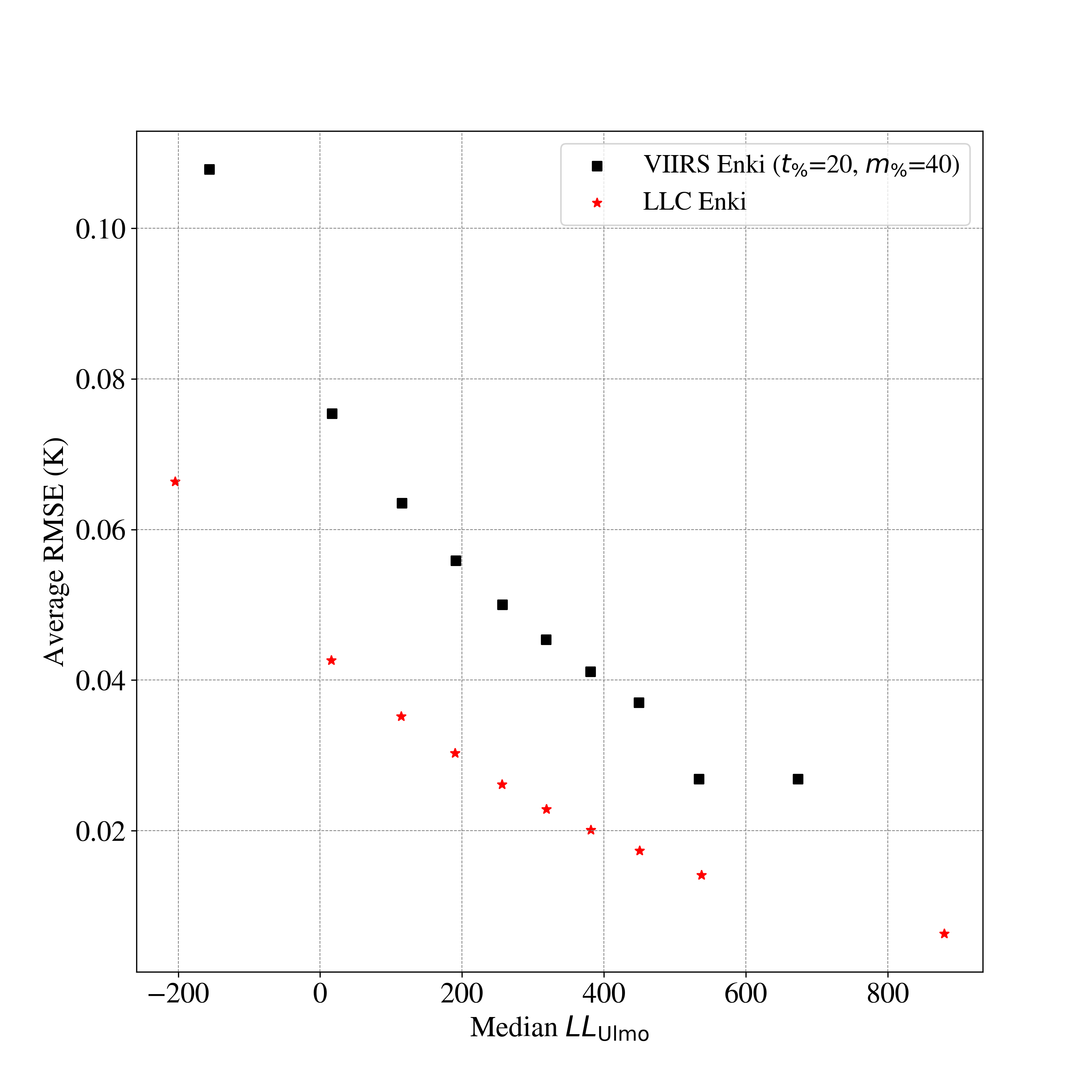}
\caption{
{\bf \enki\ performance on real sensor data.}
The black squares show the average RMSE for reconstructions of VIIRS 
images in bins of image complexity
(higher \llulmo\ indicates less complexity).
Even reconstructions of the most complex images ($\llulmo < 0$)
show
the average RMSE is
lower than the estimated sensor noise ($\approx 0.1$\,K; \cite{rs9090877}).
For comparison, we show the average RMSE values for \enki\ reconstructions
of the \llc\ validation data (red).
All of the results adopt the \enki\ model
trained with \tper=20 and applied to
cutouts with \mper=40.
}
\label{fig:viirs}
\end{figure}

As a proof of concept for reconstructing
real data, we applied {\enki} 
to the VIIRS dataset described
in Section~\ref{sec:methods}.  
For this exercise, we inserted
randomly distributed clouds into 
cloud-free data, each with a size of 
\patchsz.  Figure~\ref{fig:viirs} shows
that the average \rmse\ values for {\enki} 
are less than sensor error 
($\rmsev < \errviirs$) for cutouts with
all complexity.
The VIIRS reconstructions, however, do
show higher average RMSE than those on the
\llc\ validation dataset.
A portion of the difference is because
the latter does not include sensor noise,
which \enki\ has (sensibly) not been
trained to recreate.
We attribute additional error to the fact that
the unmasked data also suffers from sensor noise
(see Appendix~\ref{app:patches}).
And, we also anticipate a portion of the difference
is because
the VIIRS data have higher
spatial resolution than the \llc\ model.
Future experiments with higher resolution
models will test this hypothesis.

\newpage



\begin{figure}[h]%
\centering
\includegraphics[width=0.98\textwidth]{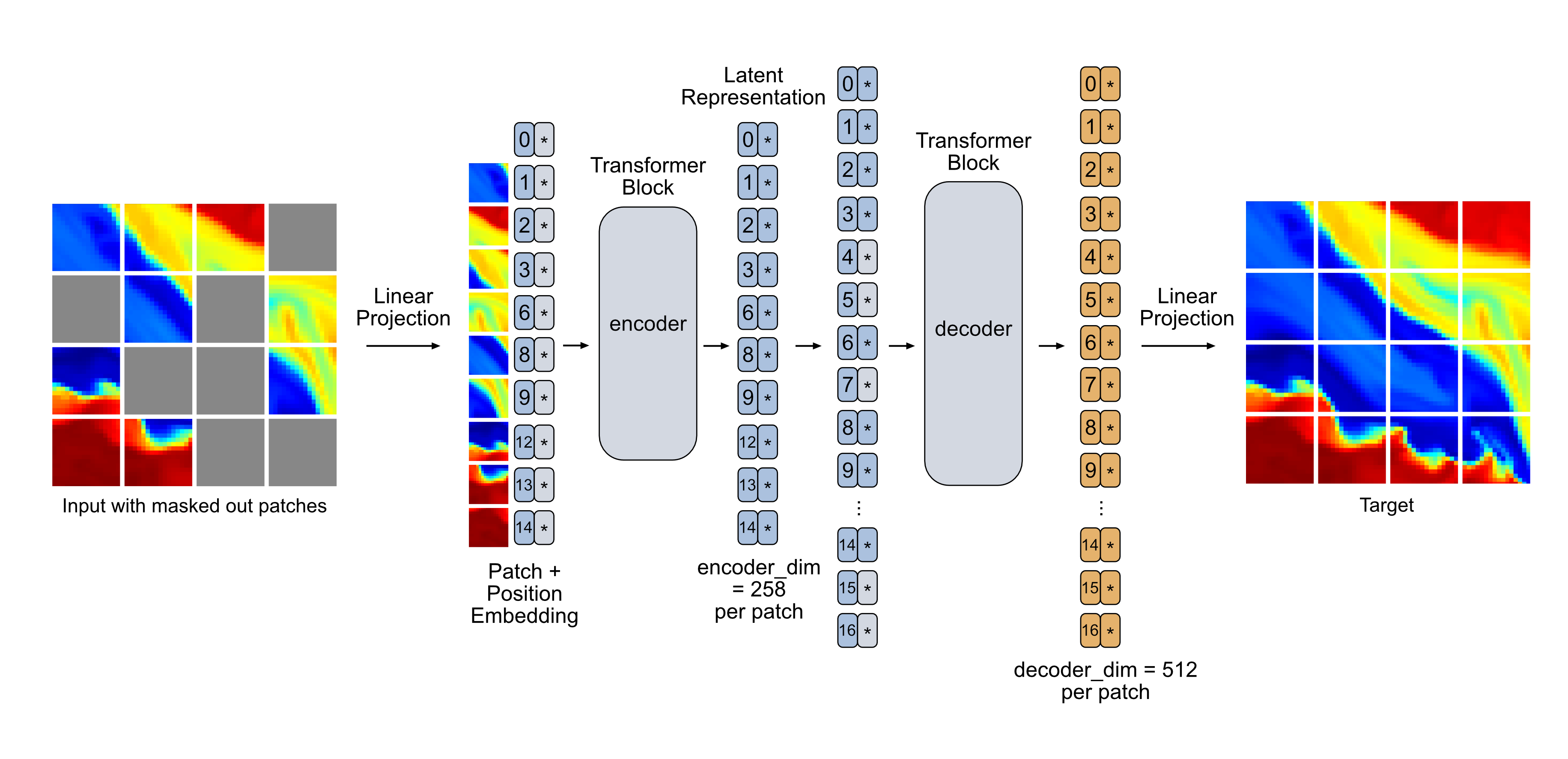}
\caption{
Architecture of our ViTMAE named \enki. In this example, the image is 64x64 pixels and broken up into patch sizes of 16x16 pixels with ~40\% of the image masked out.
An image is broken down into patches and masked. The unmasked patches are flattened and embedded by a linear projection with positional embeddings which are then run through the encoder, returning the encoded patches which are then run through the decoder along with the masked (unfilled) tokens. 
This returns another latent vector, and another linear projection layer outputs this vector as an image with the same dimensions of the original image. 
The model of \enki\ we train reconstructs cutouts of size $64\times64$\,pixels with a patch size of $4\times4$\,pixel$^2$ and a latent vector size of 256.
}
\label{fig:arch}
\end{figure}

\section{Methods}
\label{sec:methods}

We have designed and trained a machine learning model, inspired by \vitmae\ \cite{vitmae}, 
named {\enki} to reconstruct masked SST fields
\cite{aathesis}.
The architecture (Figure~\ref{fig:arch})
inputs (and outputs) a 
single-channel image, adopts a patch size of 
$4\times 4 \, \rm pixels$, 
and uses 256-dimension
latent vectors.

{\enki} was trained using SST obtained from ocean 
model output of the \llc. The {\llc} is a year-long (13 September 2011 to 14 November 2012) global ocean simulation built upon the MITgeneral circulation model \cite{Marshall:1997a,Marshall:1997b,adcroft2004cubedsphere}, \textit{i.e.}~a primitive equation model that integrates the equations of motion on a 1/48$^{\circ}$ Lat/Lon-Cap (LLC) grid. The simulation is initialized from lower resolution products of the Estimating the Circulation and Climate of the Ocean (ECCO) project and spun up at progressively higher resolutions. The model is forced at the ocean surface by European Centre for Medium-range Weather Forecasting (ECMWF) reanalysis winds, heat fluxes and precipitation \cite{dee2011era}, and by barotropic tides \cite{arbic2018tides}.

One can, of course, use `cloud-free' portions of {\sst} fields obtained from satellite-borne sensors, however, we often find undetected clouds in these fields, which 
would impact the training of \enki.
Furthermore, these fields are geographically biased \cite{ulmo}
and would yield
a highly unbalanced training set.

The {\llc} simulations have been widely used in a number of studies investigating submesoscale phenomena \cite{rocha_mesoscale_2016,su_ocean_2018,torres2018partition}, baroclinic tides \cite{savage2017spectra,arbic2018tides}, and mission support for the Surface Water, Ocean Topography (SWOT) satellite sensor \cite{wang2018swot}. As it has a horizontal grid resolution comparable to but slightly coarser than the spatial resolution of most IR satellite SST measurements, and as it is free from fine-scale atmospheric affects, it represents an oceanographic surface approximately equivalent to but reduced in noise relative to IR satellite SST. See \cite{ecco4} for further details about the implementation 
of atmospheric effects in the {\llc}. Global model-observation comparisons at these fine horizontal scales can also be found in \cite{ulmo_on_llc,yu2019llcdrifter,arbic2022drifter}.



Every two weeks beginning 2011-09-13,
we uniformly extracted \ntrain\ ``cutouts''
of $\approx 144 \times 144 \, \rm km^2$
from the global ocean at latitudes lower
than $57^\circ$N, avoiding land.
Each of these initial cutouts were re-sized to 
\cutoutsz\ with linear interpolation and
mean subtracted.  No additional pre-processing
was performed.

We constructed a complementary, validation dataset
of \nvalid~cutouts in a similar fashion.  
These were drawn from the ocean model on
the last day
of every 2 months starting 2011-09-30.
They were also offset by 0.25\,deg in longitude
from the spatial locations of the training set.

A primary hyperparameter of the \vitmae\ is the training
percentage (\tper), i.e.\ the percentage of pixels
masked during training (currently a fixed value).
While \cite{vitmae} advocates \tper=75 to insure
generalization,
we generated {\enki} models with \tper=[10,20,35,50,75].
In part, this is because we anticipated applying {\enki} 
to images with less than 50\%\ masked data ($\mper < 50$).

For the results presented here, we train using patches
with randomly assigned location (and zero overlap).
This is, however, an inaccurate representation
of actual clouds which exhibit spatial correlation
on a wide range of scales. 
Future work will explore how more representative
masking affects the results.


{\enki} was trained on eight NVIDIA-A10 GPUs on the Nautilus
computing system.
The most expensive \tper=10 model
requires 200~hours to complete 400 training epochs
with a learning rate of $lr = 10^{-4}$.
As described in \cite{aathesis}, \enki\ exhibits a systematic
bias for cases when $\mper \ll \tper$
i.e., mask fractions significantly lower than 
the training fraction.  For the results presented here,
we have calculated this bias from the validation dataset
and removed it.  For the favored model (\tper=\tbest),
however, this bias term is nearly negligible
($< 10^{-4}$\,K).

In addition to the \llc\ validation dataset,
we apply {\enki} to actual remote sensing data.
These were extracted from the level~2 product of
the NOAA processed granules of the VIIRS sensor 
\cite{viirs}.
We included data from 2012-2021 and only included
\cutoutsz\ cutouts without any masked data.
These are \nviirs~cutouts with geographic preference
to coastal regions and the equatorial Pacific 
(see \cite{ulmo_on_llc}).
We caution that while we selected `cloud-free' {\cutoutsz} regions, 
we have reason to believe that
a portion of these data are affected
by clouds (e.g.\ \cite{nenya}).


\section{Conclusion}\label{sec13}






Understanding of the Earth climate system depends heavily upon satellite measurements. In particular, a proper Earth observing system must measure quantities descriptive of the atmosphere, ocean, and land globally and at approximately daily resolution in order to better understand the interaction of these carbon reservoirs with one another. At present, such spatial and temporal coverage is afforded only by space-borne sensors. Moreover, our ability to properly predict these interactions in the future--at least in the short-term--depends strongly upon deductions made using present-day and historical satellite measurements. In summary, remotely-sensed measurements and derived quantities are critical for understanding Earth's present and future state.

The Global Climate Observing System (GCOS) is a multi-national committee or effort sponsored by the World Meteorological Organization, International Oceanographic Commission, United Nations Environment Programme, and the International Council for Science. Its chief mission is to ensure that observations and information needed to answer climate-relevant questions are obtained and made available to the public. Of the 54 {Essential Climate Variables (ECVs)} identified by GCOS, it is notable that 8 of these are variables descriptive of the ocean surface, one of these being SST. 
Moreover, ocean surface heat flux is another closely-related ECV, being heavily determined by SST. Finally, and perhaps more germane to the present discussion, of the 54 ECVs, 26 of these can be measured by satellite sensors that operate in both the visible and infrared portions of the electromagnetic spectrum, with atmospheric conditions (\textit{e.g.} aerosols, water vapour, and especially clouds) often interfering with these measurements. In fact, only $\sim 15\%$ 
of the Earth's surface is visible to these satellite sensors at any given time. It can therefore conservatively be stated that missing or masked pixels due to atmospheric conditions severely limit estimates of 26 of the 54 climate-critical variables. Clearly, these data gaps represent considerable impediments to scientific discovery and, hence, advances in climate prediction.

In this study,
we have introduced a new approach 
(i.e.~\enki) to mitigate 
missing or masked data through image reconstruction.
This dynamically-informed reconstruction technique leverages an NLP algorithm trained
on high-fidelity ocean model output to reconstruct
masked data in \sst\ fields.
We have demonstrated that the \enki\ algorithm 
has reconstruction errors less than approximately $0.1$\,K
for images with up to 50\%\ missing data. That is, the RMSE is comparable or less than typical sensor
noise.
Furthermore, \enki\ outperforms other widely-adopted 
approaches by up to an order-of-magnitude in RMSE,
especially for fields with significant SST structure.
Systems built upon \enki\ (or perhaps future algorithms like it) may therefore 
represent an optimal approach
to mitigating masked pixels in remote sensing data.

An immediate application of {\enki} is the improvement of more than 40 years of SST measurements made by polar-orbiting and geosynchronous spacecraft 
\cite{aathesis}. In addition to reduction in geographic and seasonal biases \cite{ulmo,ulmo_on_llc}, improvement of these datasets would likely translate to enhanced time series analysis and teleconnections between ECVs across the
globe. One objective of the present work, for example, is the improvement of Level 2 (\i.e.\ swath) SST from the Moderate Resolution Imaging Spectroradiometer (MODIS), a high-resolution ($1$-km pixels, twice daily) data record which extends from 2000 to the present. Additionally, we anticipate integrating portions of the {\enki} encoder within comprehensive
deep-learning models
(e.g.\ \cite{presto}) in order to 
predict dynamical processes and extrema at the ocean's surface.


While manipulation of global, realistic ocean simulations and training on these petabyte-sized datasets offer practical computational and data-storage challenges, the final algorithm is simplistic enough that we expect implementing {\enki} within an end-to-end
system to reconstruct images for 
commercial applications 
would be straightforward. 
Within this context, enhanced SST measurements would enable improved early detection and monitoring of harmful algal blooms and marine heatwaves \cite{hare2010fisheries,jacox2022nature,poloczanska2013nature}, allowing fisheries and aquaculture industries to take preventative measures and minimize economic losses. At the same time, improved analysis of \sst\ could improve weather forecasts, minimizing risk associated with severe weather events such as tropical cyclones and monsoons \cite{knutson2004hurricane,schade1999tropicalcyclone,sharmila2013miso}.


Optimal performance of {\enki} may be achieved by iterative application of models with a range of \tper\ and/or trained 
on specific geographical locations. At the minimum, improvements in the present approach will require models that accommodate a wider range of spatial scales and resolution than has been considered
here (e.g.\ \cite{scalemae}). 
This is necessary, for example, to accommodate geostationary 
\sst\ estimates, which have spatial resolutions closer to $5$-$10$~km.
We anticipate such improvements 
are straightforward to implement and are
the focus of future work. Finally, we emphasize that the work presented here may be generalized to any remote sensing datasets in which a global corpus of realistic numerical output is available. In the oceanic context, this dataset might be ocean wind vectors, sea surface salinity, ocean color and--with improved
biogeochemical modeling--even phytoplankton.


\backmatter


\bmhead{Acknowledgments}

We thank David Reiman (Deep Mind) for 
inspiring this work, and Edwin Goh for his contributions in discussions on hyperparameters and early results. We also acknowledge use of the Nautilus cloud computing system which is supported by the following US National Science Foundation (NSF) awards: CNS-1456638, CNS1730158, CNS-2100237, CNS-2120019, ACI-1540112, ACI-1541349, OAC-1826967, OAC-2112167.

%
%
%
%
%
%
%
%

\begin{appendices}

\section{Comparison with Other Approaches}\label{app:other}

\begin{figure}[h]%
\centering
\includegraphics[width=0.98\textwidth]{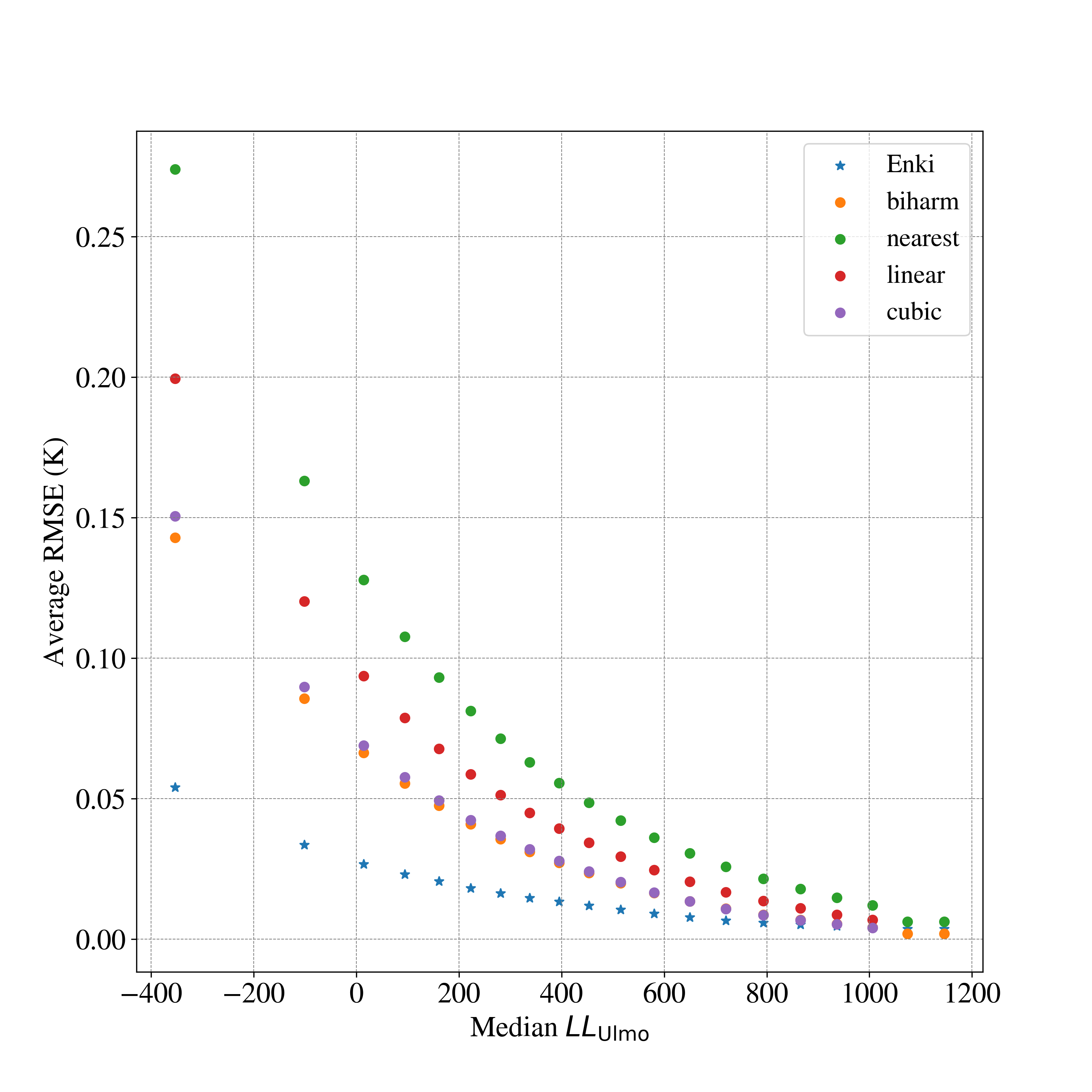}
\caption{
Comparison of the average RMSE for a series of interpolation 
schemes (circles) against the results from \enki.
These are for reconstructions of the \llc\ validation dataset
using the \tper=20 model on data \mper=30 masked patches.
Aside from the nearly featureless cutouts ($\llulmo > 800$),
\enki\ well out-performs all of these schema and
typically by factors of $3-5\times$.
}
\label{fig:many_inpaint}
\end{figure}

\cite{reconstruct_review} reviews the wide variety of approached adopted
by the community to reconstruct remote sensing data.
While a complete comparison to these is beyond the scope of this
manuscript, we present additional tests here.
Figure~\ref{fig:many_inpaint} shows the results of a series of interpolation
schemes adopted in the literature, as a function of cutout complexity
gauged by the \llulmo\ metric.
Of these, the most effective is the biharmonic inpainting algorithm
presented in the main text and adopted in our previous works \cite{ulmo}.
The figure shows, however, that \enki\ outperforms even this 
method by a factor of $2-3\times$ in average RMSE
aside from the featureless data ($\llulmo  < 300$). 

One of the most widely adopted techniques for remote
sensing reconstruction in the literature is the DINEOF
algorithm.  This approach fits an EOF to a time series
of \sst\ data to then predict masked values.
We have performed a test of the DINEOF algorithm implemented
by \cite{DINEOF2005} and provided as open-source:
https://github.com/aida-alvera/DINEOF.

\begin{figure}[h]%
\centering
\includegraphics[width=0.98\textwidth]{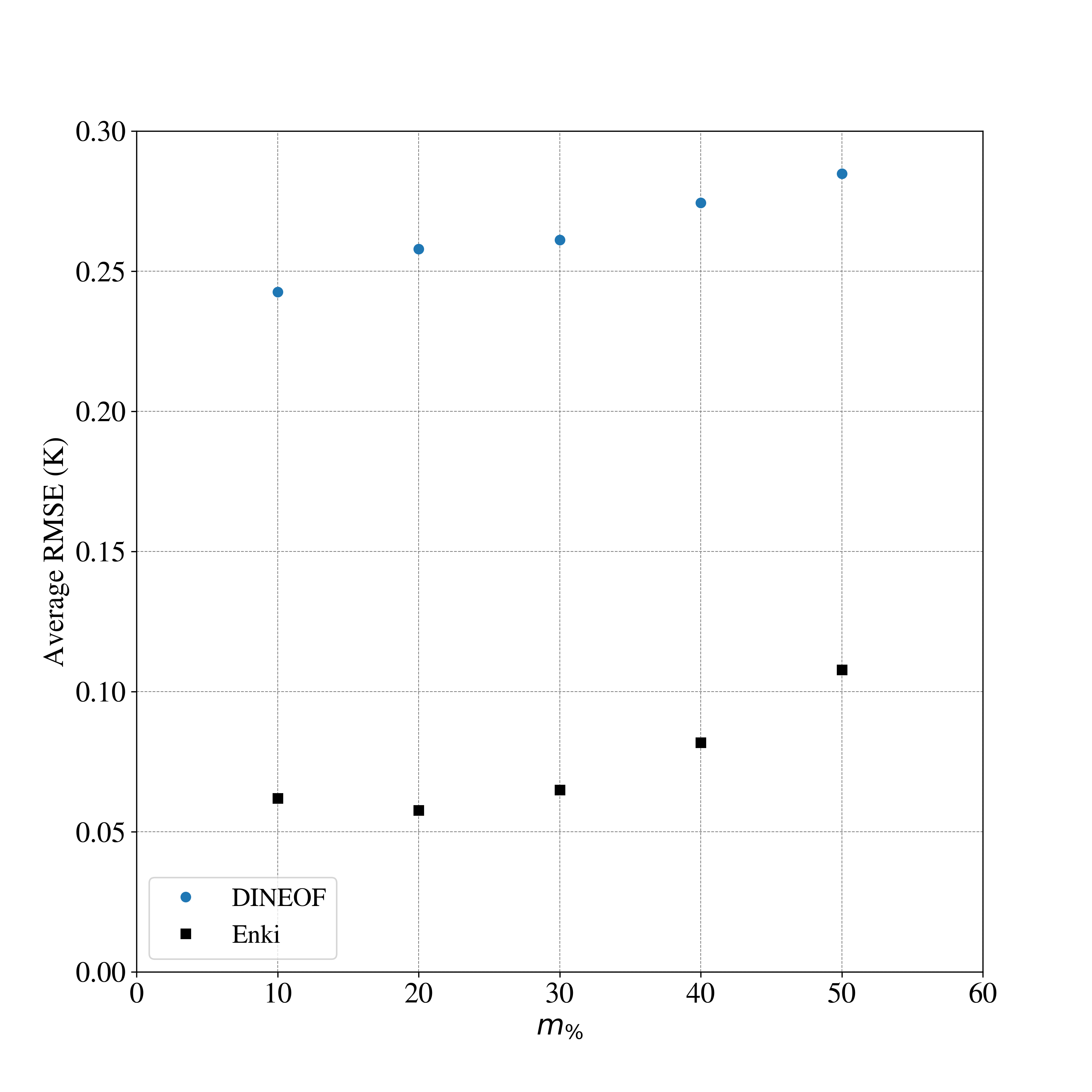}
\caption{
Average reconstruction error for the DINEOF algorithm
(blue circles) applied to 180~days of images from the China Sea.
We recover RMSE~$\approx 0.25-0.3$\,K, typical of other
published results using DINEOF.
Applying the \tper=20 \enki\ model to the same masked data,
we recover an $\approx 5\times$ lower RMSE.
We further emphasize that unlike DINEOF does not (yet)
incorporate time-series information nor was it trained
exclusively on the China Sea.
}
\label{fig:dineof}
\end{figure}

Specifically, we analyzed a 180-day sequence of cutouts 
at lat,lon=118E,21N in the China Sea.
Each of these was masked using random
$4 \times 4$~pixels patches as implemeted for \enki.
Figure~\ref{fig:dineof} presents the average RMSE 
for reconstructions with a range of masking percentiles \mper.
Similar to published results with DINEOF,
we recover an average RMSE of $\approx 0.25$\,K and
independent of \mper.
In contrast, the \enki\ reconstructions have an RMSE~$\approx 0.05$,
i.e.\ $\approx 5\times$ lower on average than the DINEOF algorithm.
We also emphasize that first results using
a convolutional neural network (DINCAE; \cite{DINCAE})
yield RMSE values similar to DINEOF.
The \vitmae\ approach of \enki\ offers a qualitative
advance over traditional deep-learning vision models.

\section{Impacts of sensor and retrieval noise on the performance of {\enki}}\label{app:patches}

To better understand the RMSE vs $\sigma_T$ trends
of Fig.~\ref{fig:patches}b and the impacts of noise 
on {\enki} reconstructions of masked areas in {\sst} fields, 
we decompose both RMSE and $\sigma_T$ into contributing components.

For satellite-derived {\sst} fields, patch complexity (herein denoted by $\sigma_T$) is a function of both (1)~noise in the patch resulting from the instrument or noise introduced as part of the retrieval process $\sigma_{noise\_o\_in}$ and (2)~geophysical structure within the field $\sigma_{geo_\_o\_in}$; i.e.,~the signal of interest in the reconstruction. In these subscripts, the nomenclature ``$\_o\_in$'' signifies that these terms relate to the original field and that the $\sigma$ values correspond to the SST field inside the patch. 
Reasons for this distinction will become clear below.

Using these definitions, 
we can express the signal variance of the patch as
\begin{equation}
\sigma_T^2 = \sigma_{noise\_o\_in}^2 + \sigma_{geo_\_o\_in}^2 + 2\sigma_{{noise\_o\_in},{geo\_o\_in}}^2 
\end{equation}
where $\sigma_{geo_\_o\_in}$ is a function of the structure of the field, which we designate as $\xi_{o\_in}$, and $\sigma_{{noise\_o\_in},{geo\_o\_in}}^2$ is the covariance between the sensor/retrieval noise and geophysical signal. 
If we assume negligible correlation between the two sources of variability, 
then we approximate

\begin{equation}
\stdT^2 \approx \sigma_{noise\_o\_in}^2 + \sigma_{geo_\_o\_in}^2,
\label{eqn:std}
\end{equation}
Moreover, the root mean squared error (RMSE) of the prediction, which constitutes our measure of the quality of the reconstructed image, is given by
\begin{equation}
	\begin{split}
 {\rm RMSE}^2(\sigma_{noise\_o\_in},\, &\xi_{o\_in},\, \sigma_{noise\_p\_in},\, \xi_{p\_in}) = \\
&\sigma_{noise\_o\_in}^2 + \sigma_{noise\_p\_in}^2 + f(\xi_{o\_in},\, \xi_{p\_in})
	\end{split}
\end{equation}
\noindent where the subscript $\_p$ references the predicted field, $\_out$ refers to the characteristic, either noise or the geophysical signal, outside of the masked areas, and $f(\xi_{o\_in},\, \xi_{p\_in})$ is the contribution to ${\rm RMSE}^2$ resulting from the difference between the geophysical structure of the original field and that of the predicted field in the masked areas. 
Because \enki\ was trained on effectively noise-free model outputs
(numerical noise will be negligible), we ignore $\sigma_{noise\_p\_in}$
hereafter.
We also emphasize that
the predicted geophysical variability $\xi_{p\_in}$ is
a function of the noise and geophysical structure in the original field  
(i.e.\, outside the masked pixels): 
\begin{equation}
	\begin{split}
        \xi_{p\_in} &= \xi_{p\_in}(\, \sigma_{noise\_o\_out},\, \xi_{o\_out}).
	\end{split}
\end{equation}
\vskip 1mm
\noindent While this complex dependence can make interpretation of 
RMSE vs $\sigma_{T}$ curves challenging, 
we offer the following discussion to aid the reader.

\begin{figure}[ht]%
\centering
\includegraphics[width=0.98\textwidth]{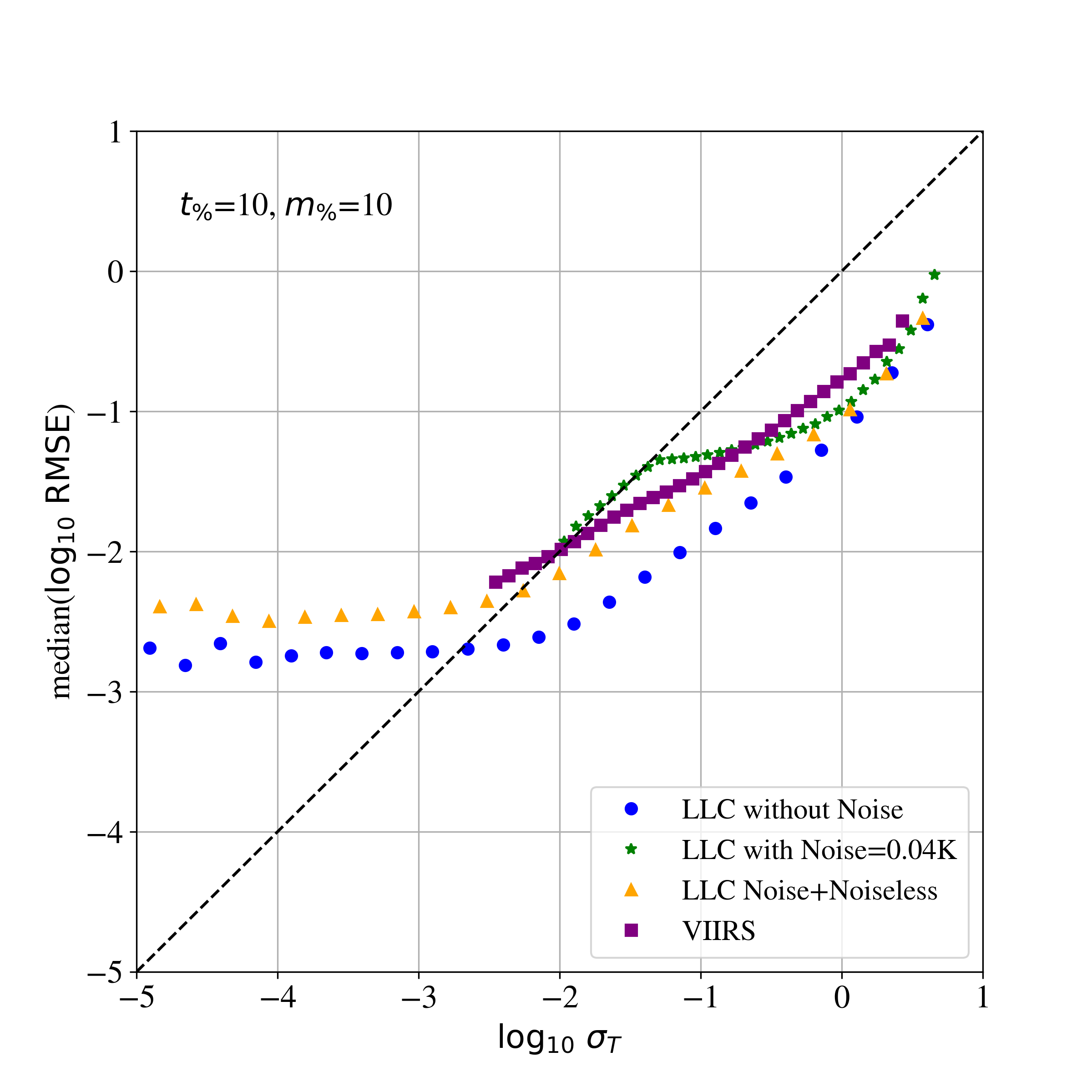}
\caption{
Investigation of the reconstruction error in individual
$4 \times 4$~pixels patches as a function of \stdT\ the measured
standard deviation within the patch.
For all of the datasets examined, we adopt the \tper=10 \enki\ model
with \mper=10 masking.
The blue circles are for the noiseless (\snoin=\snoout=0\,K)
\llc\ validation
dataset similar to Figure~\ref{fig:patches}b.
The green stars is the same dataset with but
with imputed white noise
(\snoin=\snoout=0.4\,K). 
This curve follows the one-to-one line for 
$\stdT < \snoin$ as expected and then recovers
toward the noiseless case as where the geophysical
signal dominates \stdT.
The yellow triangles, meanwhile, show results where
noise was imputed in the unmasked patches but ignored
when measuring RMSE
(\snoin=0\,K, \snoout=0.4\,K).
The difference between this case and the noiseless dataset
describe the impact of noise in the unmasked data for reconstruction.
Last, the magenta squares show results for the VIIRS dataset.
See the text for additional discussion on those results.
}
\label{fig:many_patches}
\end{figure}

The blue markers in Fig~\ref{fig:many_patches} 
correspond to the noise-free case (i.e., $\sigma_{noise\_o\_in} = \sigma_{noise\_o\_out} = 0$):
\begin{equation}\label{noisefree/noisefree}
    {\rm RMSE}^2 = 
       f(\xi_{o\_in},\, \xi_{p\_in}(\, 0,\, \xi_{o\_out})).
\end{equation}
\vskip 1mm
\noindent In the present context, this corresponds to the case where the original field is model SST. We attribute the flat portion of the curve up to $\sigma_T\approx 10^{-2}$ to two effects.  
The first is the limited precision of the \vitmae\ tokenization of
the patches;  here we have adopted 256\,dimension latent vector.
The other effect is
a correlation between the size of geophysical structures and the magnitude of $\sigma_T$. 
Specifically, for {\stdT}$\lesssim 0.02$\,K the spatial scale of oceanographic features is on the order of or smaller than the $4\times 4$ pixel patch size (i.e.~$8\times8$\,km$^2$), such that there is little to no information available to reconstruct the structure in the masked area. As {\stdT} increases above these values, the spatial scale of the features increases, with more information in the surrounding field available to reconstruct the field within the patch.

But the above analysis and interpretation is free from noise typically encountered in satellite-derived measurements. As previously mentioned, this can occur either due sensor noise or errors due to limitations of the retrieval algorithms used to estimate {\sst} from the measured radiance. To investigate the impact of sensor and retrieval noise on {\enki}'s reconstructions, we add 0.04\,K white Gaussian noise to the {\llc} cutouts ($\sigma_{noise\_o\_in} = \sigma_{noise\_o\_out} = 0.04$~K in the above nomenclature) giving:
\begin{equation}\label{noise/noise}
    {\rm RMSE}^2 = (0.04~\mathrm{K})^2 + 
      f(\xi_{o\_in},\, \xi_{p\_in}(\, 0.04~\mathrm{K},\, \xi_{o\_out})) \;\;.
\end{equation}
\noindent We then repeat the analysis. Not surprisingly, the new results (green stars in Fig.~\ref{fig:many_patches}) follow the 1:1 line up to $\sigma_T$ of approximately 0.04\,K, after which the RMS difference between the masked portions of the reconstructed {\sst} fields and the underlying {\sst} fields, to which 0.04\,K Gaussian noise was also added, becomes progressively smaller than $\sigma_T$. For $\sigma_T\lessapprox0.04$\,K, the added noise tends to obscure the geophysical structure of the field. As $\sigma_T$ increases, however, the structure in the field eventually overwhelms the added noise and the improvement in reconstruction approaches that achieved with no noise added--i.e.~the blue circles in Fig.~\ref{fig:many_patches}.

Also shown in Fig.~\ref{fig:many_patches} is a similar set of points for VIIRS cutouts, the magenta squares. 
This curve follows neither the {\llc} curve without noise (blue circles) nor the {\llc} curve with noise (green stars). We believe that this results from the non-Gaussian nature of the noise in the VIIRS cutouts. To explore this, we repeat the above analysis except for $\sigma_{noise\_o\_in} = 0$ and $\sigma_{noise\_o\_out} = 0.04~\mathrm{K}$; i.e.,
\begin{equation}\label{noise/noisefree}
    {\rm RMSE}^2 = \sigma_{noise\_p\_in}^2(\, 0.04~\mathrm{K},\, \xi_{o\_out}) + f(\xi_{o\_in},\, \xi_{p\_in}(\, 0.04~\mathrm{K},\, \xi_{o\_out})),
\end{equation}
\noindent which results in the yellow triangles in Fig.~\ref{fig:many_patches}. The added noise in this case only contributes to RMSE via {\enki}'s reconstructed fields; the difference between the blue circles
(Eq.~\ref{noisefree/noisefree}) and yellow triangles (Eq.~\ref{noise/noisefree}) is a measure of the impact of the noise added to the region outside of masked areas on {\enki}'s reconstructed fields in masked areas. 
This curve is more similar to the VIIRS curve than the curves for either of the other two cases, no-noise (Eq.~\ref{noisefree/noisefree}) or Gaussian noise (Eq.~\ref{noise/noise}), for $10^{-3}\lessapprox\sigma_T\lessapprox2\times10^{-1}$\,K, the range including in excess of 80\% of cutouts. This suggests that the noise in the VIIRS fields is not Gaussian. The two primary contributors to non-Gaussian VIIRS noise are (1) {\it Instrument noise} (VIIRS is a multi-detector instrument and this can introduce noise in the along-track direction at harmonics corresponding to the number of detectors) and (2) {\it Clouds that were not properly masked by the retrieval algorithm.} Clouds that have not been properly masked tend to result in cold anomalies, often substantially colder than the surrounding cloud-free region, which are structurally incompatible with geophysical processes. Furthermore, such anomalies tend to be relatively small in area--5 to 20 pixels--and in number\footnote{We have been surprised by the significant fraction of cutouts in the VIIRS, `cloud-free' product we are using that are affected in this fashion.}. Although both of these sources of non-Gaussian noise may contribute to the shape of the VIIRS RMSE vs $\sigma_T$ curve we believe that the primary problem is related to improperly masked clouds or other small scale atmospheric phenomena, which imprint themselves on the {\sst} field as part of the retrieval. 

In the above, we have shown how noise in the area surrounding masked pixels affects {\enki}'s ability to reconstruct the masked portion of the field. While not a major focus of this work, we have also suggested that non-Gaussian noise in VIIRS {\sst} fields due to clouds, which have not been properly masked, is likely the primary cause of the degradation in {\enki}'s ability to reconstruct masked portions of these data.





\end{appendices}


\bibliography{sn-bibliography}

\end{document}